\documentclass[11pt,a4paper]{article}
\usepackage{amsmath}
\usepackage[mathcal]{eucal}
\usepackage{latexsym}
\usepackage{amssymb}
\begin{titlepage}
\title{Instanton representanton of Plebanski gravity. Gravitational coherent states}
\author{Eyo Eyo Ita III}
\medskip
\input amssym.def
\input amssym.tex

\def \in{\indent}

\begin{document}
\maketitle
\bigskip
\centerline{Department of Applied Mathematics and Theoretical Physics} 
\smallskip
\centerline{Centre for Mathematical Sciences, University of Cambridge, Wilberforce Road}
\smallskip
\centerline{Cambridge CB3 0WA, United Kingdom}
\smallskip
\centerline{eei20@cam.ac.uk} 

\bigskip

\begin{abstract}
In this paper we show that the instanton representation of Plebanski gravity exhibits a Hilbert space of harmonic oscillator-like coherent states.  We put in place the formalism and carry out the construction of the states, and we elucidate on their physical interpretation.  Additionally, we provide an invertible map between the Ashtekar variables and this Hilbert space of states, via the instanton representation.  Finally, we compare and constrast our formalism and some of our results with the corresponding ones in loop quantum gravity.
\end{abstract}
\end{titlepage}

\section{Introduction}

\noindent
Loop quantum gravity (LQG) is a background-independent, nonperturbative approach to the canonical quantization of gravity.\footnote{For a well-written exposition of LQG, the reader is directed to \cite{LOOP} and \cite{LOOP1}.  These will provide a sufficent level of review for our purposes.  For greater depth into LQG, the associated references therein should provide a fairly comprehensive account.}  In this approach one wishes to construct a physical Hilbert space of states solving the initial value constraints of general relativity (GR), using a kinematical Hilbert 
space $H_{Kin}$ as the starting point.  The spin network states are defined on one-dimensional graphs embedded in 3-space, and form an orthonormal basis for $H_{Kin}$.  These states are eigenstates of area and have been useful for the understanding of gauge invariance at the quantum level, and for verification of the Bekenstein--Hawking law for black hole entropy.  Due to nonseperability of $H_{Kin}$, the interpretation of infinitesimal diffeomorphisms becomes problematic in LQG.  A diffeomorphism-invariant Hilbert 
space $H_{diff}$ can be obtained by group averaging of states in $H_{Kin}$, whereupon one attempts to find the physical Hilbert space $H_{Phys}$ implementing the Hamiltonian dynamics.  The quantization of the Hamiltonian constraint in LQG is subject to various ambiguities, the meaning of which are unclear, and is currently still a matter of debate.  Additionally, the existence of states satisfying the constraints which correlate in the classical limit to a well-defined geometry, is as well an outstanding issue.  The spin foam formalism can be seen as a covariant form of LQG which attempts to implement the dynamics of gravity via the Plebanski action using path integrals.\par
\indent  
In this paper we will attempt to achieve some of the aims of LQG using a new approach which we have called the instanton representation of Plebanski gravity.  In this approach is implicit a natural algebra of observables for what we will regard as the physical degrees of freedom for certain sectors of GR.  We will show that the representation of this algebra exhibits a natural coherent state structure for gravity, as well as a natural quantization of the physical degrees of freedom.  First we will implement the quantum Hamiltonian dynamics on the kinematic phase space $\Omega_{Kin}=(\Gamma_{Kin},P_{Kin})$, namely the phase space variables of the theory at the level after implementation of the diffeomorphism and the Gauss' law constraints, and prior to the Hamiltonian constraint.\footnote{For notational purposes, $\Gamma_{Kin}$ will refer to configuration space at this level and $P_{Kin}$ will refer to the corresponding momentum space}  Then we will focus on the (inverse) projection from the full unconstrained phase space to $\Omega_{Kin}$, and the corresponding map to the Ashtekar variables.  Note that the Ashtekar variables comprise the starting point for application of the loop quantization programme.\par
\indent  
The basic momentum space variables in our new approach are the densitized eigenvalues of the antiself-dual Weyl curvature (CDJ matrix), which will constitute the physical degrees of freedom.  The Hilbert space of the theory will be defined with respect to $\Omega_{Kin}$, where these degrees of freedom are explicit.  It is on this space where the quantization procedure and the implementation of the reality conditions have been defined.  One feature of the Hilbert space as constructed in \cite{EYO} is that the states resemble an infinite-dimensional analogue of harmonic oscillator-like coherent states, which are applicable for vanishing cosmological 
constant $\Lambda$.  In \cite{ITAE} we have generalized the construction to include nonvanishing $\Lambda$, which entails the use of holomorphic hypergeometric functions.\footnote{In \cite{EYO1} we have treated the implementation of reality conditions at the kinematic level both for $\Lambda=0$ and for $\Lambda\neq{0}$, including via adjointness relations on the Hilbert space.}  In the present paper we will carry out the construction of the states within the context of the coherent state formalism, and will demonstrate that these states are annihilated the Hamiltonian constraint.\par
\indent
The organization of this paper follows a bottom-up rather than the conventional top-down approach, as we will first establish a system of coherent states and then provide a map from this system to the Ashtekar variables, via the instanton representation.  The organization of this paper is as follows.  In section 2 we provide a brief review of the oscillator formalism and coherent states, building on the relevant concepts from \cite{COHERENT} 
and \cite{PERELOMOV}.  The purpose of this section is to put in place the formalism, and to introduce the constituents of some of the operators which will have direct analogues for gravity.  Sections 3, 4 and 5 carry out the transformation from the coherent state basis and operators into the holomorphic Schr\"odinger representation, from which we derive the Ashtekar variables via the instanton representation.  We have also outlined the solution to the Hamiltonian constraint in the holomorphic Schr\"odinger representation in terms of hypergeometric functions.  The association of the gravitational Hilbert space with oscillator coherent states uniquely picks out the Bargmann representation \cite{BARGMANN} and the accompanying adjointness relations.  In section 6 we provide a brief physical interpretation of the states and what features of spacetime they describe.  In section 7 we outline the construction of the hypergeometric solutions to the Hamiltonian constraint using a Lippman--Schwinger type expansion with respect to the coherent state basis.  In this section we formalize the link from the coherent states to the gravitational degrees of freedom using the holomorphic Schr\"odinger representation.  Section 8 provides a brief discussion of our results in relation to spin foams and LQG.

\section{Quantum harmonic oscillator formalism}

We will first start with a simple system, where all of the the steps of the algebraic extension to Dirac's quantization procedure, outlined in \cite{ALGEBRAIC}, can be carried out to completion.  Our system consists of three uncoupled simple harmonic oscillators with annihilation operators $a_1$, $a_2$ and $a_3$.  From $a_f$ construct the following set $S$, given by

\begin{eqnarray}
\label{WEHAVE}
S=\Bigl{\{}a_1,a_2,a_3,a^{*}_1,a^{*}_2,a^{*}_3,1\Bigr{\}}.
\end{eqnarray}

\noindent
It is clear from (\ref{WEHAVE}) that $S$ is closed under complex conjugation.  Additionally, $S$ is closed under the Poisson bracket since as one can easily verify from the harmonic oscillator algebra, 

\begin{eqnarray}
\label{WEHAVE1}
\{a_f,a^{*}_g\}=\delta_{fg};~~\{a_f,a_g\}=\{a^{*}_f,a^{*}_g\}=\{a_f,1\}=\{a^{*}_f,1\}=0.
\end{eqnarray}

\noindent
From (\ref{WEHAVE1}) the objects $a_f$ and $a^{*}_f$ may be regarded as the fundamental dynamical variables of a phase space $\Omega_{Kin}$.  Define $F$ as the set of all suitably regular functions on $\Omega_{Kin}$ which can be obtained as a sum of products of elements $F^{(i)}\in{S}$.  Some examples of elements of $F$ are given by\footnote{These particular functions will take on the interpretation as $SO(3,C)$ invariants which appear in the Hamiltonian constraint for gravity.}

\begin{eqnarray}
\label{WEHAVE2}
{Q}={a}_3{a}_3+{2 \over 3}({a}_1+{a}_2){a}_3+{1 \over 3}{a}_1{a}_2;\nonumber\\
{O}={a}_3({a}_3+{a}_1)({a}_3+{a}_2);~~{\tau}={a}_3+{1 \over 3}({a}_1+{a}_2).
\end{eqnarray}

\noindent
Next, we will associate with each element $F^{(i)}$ in $S$ an abstract operator $\hat{F}^{(i)}$, and construct the free algebra $\boldsymbol{A}$ generated by these elementary quantum operators.  This amounts to the 
promotion of (\ref{WEHAVE}) to

\begin{eqnarray}
\label{WEHAVE3}
\boldsymbol{A}=\Bigl{\{}\hat{a}_1,\hat{a}_2,\hat{a}_3,\hat{a}^{\dagger}_1,\hat{a}^{\dagger}_2,\hat{a}^{\dagger}_3,\hat{1}\Bigr{\}},
\end{eqnarray}

\noindent
whence the Poisson brackets (\ref{WEHAVE1}) become promoted to commutators

\begin{eqnarray}
\label{WEHAVE4}
\bigl[\hat{a}_f,\hat{a}^{\dagger}_g]=\delta_{fg};~~\bigl[\hat{a}_f,\hat{a}_g]=\bigl[\hat{a}^{\dagger}_f,\hat{a}^{\dagger}_g]=\bigl[\hat{a}_f,\hat{1}]=\bigl[\hat{a}^{\dagger}_f,\hat{1}]=0.
\end{eqnarray}

\noindent
Note that (\ref{WEHAVE4}) can also be derived by application of an involution operation to (\ref{WEHAVE}).  Additionally, the promotion $S\rightarrow\boldsymbol{A}$ extends to the set $\{F\}$, hence (\ref{WEHAVE2}) become promoted to

\begin{eqnarray}
\label{DEFINING}
\hat{Q}=\hat{a}_3\hat{a}_3+{2 \over 3}(\hat{a}_1+\hat{a}_2)\hat{a}_3+{1 \over 3}\hat{a}_1\hat{a}_2;\nonumber\\
\hat{O}=\hat{a}_3(\hat{a}_3+\hat{a}_1)(\hat{a}_3+\hat{a}_2);~~\hat{\tau}=\hat{a}_3+{1 \over 3}(\hat{a}_1+\hat{a}_2)
\end{eqnarray}

\noindent
with adjoints

\begin{eqnarray}
\label{DEFININGIT}
\hat{Q}^{\dagger}=\hat{a}^{\dagger}_3\hat{a}^{\dagger}_3+{2 \over 3}(\hat{a}^{\dagger}_1+\hat{a}^{\dagger}_2)\hat{a}^{\dagger}_3+{1 \over 3}\hat{a}^{\dagger}_1\hat{a}^{\dagger}_2;\nonumber\\
\hat{O}^{\dagger}=\hat{a}_3^{\dagger}(\hat{a}_3^{\dagger}+\hat{a}_1^{\dagger})(\hat{a}^{\dagger}_3+\hat{a}^{\dagger}_2);~~\hat{\tau}^{\dagger}=\hat{a}^{\dagger}_3+{1 \over 3}(\hat{a}^{\dagger}_1+\hat{a}^{\dagger}_2)
\end{eqnarray}

\noindent
under the involution operation.  As an aside, these operators satisfy the algebra

\begin{eqnarray}
\label{DEFINING1}
\bigl[\hat{a}_3,\hat{O}^{\dagger}\bigr]=3\hat{Q}^{\dagger};~~\bigl[\hat{a}_3,\hat{Q}^{\dagger}\bigr]=2\hat{\tau}^{\dagger};~~\bigl[\hat{a}_3,\hat{\tau}^{\dagger}\bigr]=1;\nonumber\\
\bigl[\hat{O},\hat{Q}\bigr]=\bigl[\hat{Q},\hat{\tau}\bigr]=\bigl[\hat{\tau},\hat{O}\bigr]=0.
\end{eqnarray}

\par
\indent
We will now construct a linear representation of the abstract algebra $\boldsymbol{A}$ given by (\ref{WEHAVE3}).  Along with the relations (\ref{WEHAVE4}) come a unique normalized 
ground state $\bigl\vert{0},0,0\bigr>=\bigl\vert{0}\bigr>\otimes\bigl\vert{0}\bigr>\otimes\bigl\vert{0}\bigr>$ with $\bigl<0,0,0\bigl\vert{0},0,0\bigr>=1$, such that 

\begin{eqnarray}
\label{GROUND1}
\hat{a}_f\bigl\vert{0},0,0\bigr>=\bigl<0,0,0\bigr\vert\hat{a}_f^{\dagger}=0,
\end{eqnarray}

\noindent
where the creation operator in (\ref{GROUND1}) acts to the left on the bra state.  Also, we have that

\begin{eqnarray}
\label{GROUND2}
\hat{a}^{\dagger}_1\bigl\vert{0},0,0\bigr>=\bigl\vert{1},0,0\bigr>;~~
\hat{a}^{\dagger}_2\bigl\vert{0},0,0\bigr>=\bigl\vert{0},1,0\bigr>;~~
\hat{a}^{\dagger}_3\bigl\vert{0},0,0\bigr>=\bigl\vert{0},0,1\bigr>,
\end{eqnarray}

\noindent
such that for an arbitrary state $\bigl\vert{p},q,s\bigr>$ with $p\geq{0}$, $q\geq{0}$ and $r\geq{0}$,\footnote{We require for all states that $\bigl\vert{p},r,s\bigr>=0$ for any of $p$, $r$, $s$ less than zero.}

\begin{eqnarray}
\label{GROUND3}
C_{l,m,n}^{l^{\prime}m^{\prime}n^{\prime}}(\hat{a}_1)^l(\hat{a}_2)^m(\hat{a}_3)^n(\hat{a}^{\dagger}_1)^{l^{\prime}}(\hat{a}^{\dagger}_2)^{m^{\prime}}(\hat{a}^{\dagger}_3)^{n^{\prime}}\bigl\vert{p},q,s\bigr>\nonumber\\
=\bigl\vert{p-l+l^{\prime}},q-m+m^{\prime},s-n+n^{\prime}\bigr>
\end{eqnarray}

\noindent
where we have defined

\begin{eqnarray}
\label{GROUND31}
C_{l,m,n}^{l^{\prime}m^{\prime}n^{\prime}}=\sqrt{{{(l^{\prime})!(m^{\prime})!(n^{\prime})!} \over {(l!)(m!)(n!)}}}.
\end{eqnarray}

\noindent
These states form a mode number basis satisfying orthogonality relations

\begin{eqnarray}
\label{GROUND4}
\bigl<l,m,n\bigl\vert{p},q,s\bigr>=\delta_{lp}\delta_{mq}\delta_{ns},
\end{eqnarray}

\noindent
and $\hat{a}_f$ and $\hat{a}_f^{\dagger}$ are adjoints with respect to the inner product (\ref{GROUND4}).\par
\indent

\subsection{Coherent states}

In this paper we will rather be utilizing a basis of coherent states, applying the formalism of \cite{COHERENT} to our model for gravity.  One way to define coherent states is states which are eigenstates of the annihilation 
operators $a_1$, $a_2$ and $a_3$, such that for states $\bigl\vert\alpha,\beta,\lambda\bigr>=\bigl\vert\alpha\bigr>\otimes\bigl\vert\beta\bigr>\otimes\bigl\vert\lambda\bigr>$, 

\begin{eqnarray}
\label{REPRESENT28}
\hat{a}_1\bigl\vert\alpha,\beta,\lambda\bigr>=\alpha\bigl\vert\alpha,\beta,\lambda\bigr>;~~
\hat{a}_2\bigl\vert\alpha,\beta,\lambda\bigr>=\beta\bigl\vert\alpha,\beta,\lambda\bigr>;~~
\hat{a}_3\bigl\vert\alpha,\beta,\lambda\bigr>=\lambda\bigl\vert\alpha,\beta,\lambda\bigr>.
\end{eqnarray}

\noindent
We will distinguish $a_3$ as being special in relation to $a_1$ and $a_2$, since the operators of interest (for example as in (\ref{DEFINING})) will be invariant under interchange of $a_1$ and $a_2$ but not with respect to $a_3$.  Hence in terms of the individual states in the direct product, we have the definitions

\begin{eqnarray}
\label{WEHAVETHE}
\bigl\vert\alpha\bigr>=e^{-\vert\alpha\vert^2}e^{\alpha{a}_1^{\dagger}}\bigl\vert{0}\bigr>;~~
\bigl\vert\beta\bigr>=e^{-\vert\beta\vert^2}e^{\beta{a}_2^{\dagger}}\bigl\vert{0}\bigr>;~~
\bigl\vert\lambda\bigr>=e^{\alpha{a}_3^{\dagger}}\bigl\vert{0}\bigr>
\end{eqnarray}

\noindent
where $\alpha$, $\beta$ and $\lambda$ are dimensionless state labels.  It will be convenient to define the coherent states, in the sense of Perelemov \cite{PERELOMOV}, as the states obtained by application of a displacement operator to the vacuum state $\bigl\vert{0},0,0\bigr>$, or any appropriate fiducial state.  This is given by

\begin{eqnarray}
\label{REPRESENT26}
\bigl\vert\alpha,\beta,\lambda\bigr>=D(\alpha,\beta,\lambda)\bigl\vert{0},0,0\bigr>,
\end{eqnarray}

\noindent
where we have defined the displacement operator

\begin{eqnarray}
\label{REPRESENT261}
D(\alpha,\beta,\lambda)=e^{\alpha{a}_1^{\dagger}-\alpha^{*}a_1}
e^{\beta{a}_2^{\dagger}-\beta^{*}a_2}e^{\lambda{a}_3^{\dagger}}.
\end{eqnarray}

\noindent
The coherent states are obtained by displacing the vacuum state into $C_3$, a 3-dimensional complex manifold representing three copies of the complex plane.  $C_3$ at the present level plays the role of the coset space for the group manifold of three copies of the complexified Heisenberg algebra $({H}_4)^3$.\footnote{We have taken for granted the quotienting of the Heisenberg group $H_4\otimes{H}_4\otimes{H}_4$ by the identity group element $1$ and the mode 
number operator $\hat{N}_f=\hat{a}_f^{\dagger}\hat{a}_f$ in our notation, which yields the coset space $({H}_4)^3/(U(1))^6$.}  Hence (\ref{REPRESENT261}) is a typical representative in this coset space and there is a one-to-one correspondence between states $\bigl\vert\alpha,\beta,\lambda\bigr>$ and points in $C_3$.\par
\indent
There exists a natural Euclidean metric on $C_3$ which can be used to define the distance between two states labelled by $z$ and $z^{\prime}$, given by

\begin{eqnarray}
\label{NATURALMETRIC}
d(z,z^{\prime})={1 \over 2}\Bigl[\vert\alpha-\alpha^{\prime}\vert^2+\vert\beta-\beta^{\prime}\vert^2+\vert\lambda-\lambda^{\prime}\vert^2\Bigr].
\end{eqnarray}

\noindent
This metric induces the following overlap between coherent states

\begin{eqnarray}
\label{NATURALMETRIC1}
\bigl\vert\bigl<z\bigl\vert{z}^{\prime}\bigr>\bigr\vert^2=e^{-d(z,z^{\prime})}.
\end{eqnarray}

\noindent
However, for the purpose of the gravitational coherent states we will omit the last term of (\ref{NATURALMETRIC}), since we will impose a constraint which reduces $\lambda\rightarrow\lambda_{\alpha,\beta}$ to a function 
of just $\alpha$ and $\beta$.  Hence $(\alpha,\beta)\in{C}_2$ will coordinatize the physical degrees of freedom, which makes normalization with respect to the $\lambda$ label redundant.\footnote{In particular, we will associate $a_3$ and $a_3^{\dagger}$ with a time variable on the kinematic phase space $\Omega_{Kin}$, and one does not normalize a wavefunction in time.}  Therefore the overlap between two states for our purposes will involve only the $\alpha$ 
and $\beta$ labels, given by

\begin{eqnarray}
\label{REPRESENT27}
\bigl\vert\bigl<\alpha,\beta\bigl\vert\alpha^{\prime},\beta^{\prime}\bigr>\bigr\vert^2=e^{-\vert\alpha-\alpha^{\prime}\vert^2}e^{-\vert\beta-\beta^{\prime}\vert^2},
\end{eqnarray}

\noindent
and we will from now on omit $\lambda$ from the labels in the anticipation of implementing the aformentioned constraint.\par
\indent
We will be using the following resolution of the identity for the states

\begin{eqnarray}
\label{RESOLUTION}
\int{{d^2\alpha{d}^2\beta} \over {\pi^2}}\bigl\vert\alpha,\beta\bigr>\bigl<\alpha,\beta\bigr\vert=I.
\end{eqnarray}

\noindent
Note, since the states are labelled by continuous indices in a Hilbert space that has a countable basis, they are overcomplete.  Note that any arbitrary state $\bigl\vert\boldsymbol{\psi}\bigr>$ can be expanded in terms of these coherent states \cite{COHERENT}

\begin{eqnarray}
\label{CANBE}
\bigl\vert\boldsymbol{\psi}\bigr>=\int\bigl\vert\alpha,\beta\bigr>{f}(\alpha^{*},\beta^{*})e^{-\vert\alpha\vert^2/2}e^{-\vert\beta\vert^2/2}{{d^2\alpha{d}^2\beta} \over {\pi^2}},
\end{eqnarray}

\noindent
where the analytical function $f(\alpha^{*},\beta^{*})$, the coherent state representation of $\bigl\vert\boldsymbol{\psi}\bigr>$ is given by

\begin{eqnarray}
\label{CANBE1}
f(\alpha^{*},\beta^{*})=\bigl<\alpha,\beta\bigl\vert\boldsymbol{\psi}\bigr>e^{\vert\alpha\vert^2/2}e^{\vert\beta\vert^2/2}=\sum_{m,n}c_{m,n}{{(\alpha^{*})^m(\beta^{*})^n} \over {(m!)^{1/2}(n!)^{1/2}}}
\end{eqnarray}

\noindent
with $c_{m,n}$ the mode basis expansion coefficients in the expansion

\begin{eqnarray}
\label{CANBE2}
\bigl\vert\boldsymbol{\psi}\bigr>=\sum_{m,n}c_{m,n}\bigl\vert{m},n\bigr>=\sum_{m,n}c_{m,n}{{(a_1^{\dagger})^m(a_2^{\dagger})^n} \over {(m!)^{1/2}(n!)^{1/2}}}\bigl\vert{0},0\bigr>.
\end{eqnarray}

\subsection{Action of the constituent operators}

\noindent
We will now put in place the constitutents of the operator which we will use to impose constraints on our system to reduce the coset state manifold 
from $\bigl\vert\alpha,\beta,\lambda\bigr>\sim{C}_3$ to $\bigl\vert\alpha,\beta\bigr>\sim{C}_2$.  The operators $\hat{O}$, $\hat{Q}$ and $\hat{\tau}$ from (\ref{DEFININGIT}) have the following action on the coherent states

\begin{eqnarray}
\label{ACTIONON}
\hat{Q}\bigl\vert\alpha,\beta,\lambda\bigr>
=(\lambda+\gamma^{-})(\lambda+\gamma^{+})\bigl\vert\alpha,\beta,\lambda\bigr>,
\end{eqnarray}

\noindent
where we have defined

\begin{eqnarray}
\label{ACTIONON1}
\gamma^{\pm}={1 \over 3}\Bigl(\alpha+\beta\pm\sqrt{\alpha^2-\alpha\beta+\beta^2}\Bigr)\equiv\lambda_{\alpha,\beta}
\end{eqnarray}

\noindent 
as the roots of $Q$ from (\ref{WEHAVE2}), regarded as a quadratic polynomial in $a_3$.  Also we have the following actions

\begin{eqnarray}
\label{ACTIONON}
\hat{O}\bigl\vert\alpha,\beta,\lambda\bigr>
=\lambda(\lambda+\alpha)(\lambda+\beta)\bigl\vert\alpha,\beta,\lambda\bigr>;\nonumber\\
\hat{\tau}\bigl\vert\alpha,\beta,\lambda\bigr>
=\bigl(\alpha+\beta+{1 \over 3}\lambda\bigr)\bigl\vert\alpha,\beta,\lambda\bigr>.
\end{eqnarray}

\noindent
From these operators construct the following Hamiltonian constraint operators for our theory, given by

\begin{eqnarray}
\label{DEFINING2}
\hat{H}_1=\hat{Q}+l\hat{O}e^{-a_3^{\dagger}};~~\hat{H}_2=\hat{O}+r\hat{Q}e^{a_3^{\dagger}};~~l={1 \over r}
\end{eqnarray}

\noindent
where $r\neq{0}$ is a numerical constant.  The aim of this paper will be to construct states annihilated by $\hat{H}_1$ and $\hat{H}_2$ using the coherent state basis.  Part of this process will utilize the coherent states annihilated by $\hat{Q}$ and $\hat{O}$.  These are

\begin{eqnarray}
\label{ACTIONON2}
\bigl\vert\alpha,\beta,\lambda_{\alpha,\beta}\bigr>\in{Ker}\{\hat{Q}\}
\end{eqnarray}

\noindent
with $\lambda=\lambda_{\alpha,\beta}$ given by (\ref{ACTIONON1}), and

\begin{eqnarray}
\label{ACTIONON3}
\bigl\vert\alpha,\beta,0\bigr>,\bigl\vert\alpha,\beta,-\alpha\bigr>,\bigl\vert\alpha,\beta,-\beta\bigr>\in{Ker}\{\hat{O}\}.
\end{eqnarray}

\noindent
For those states annihilated by $\hat{H}_1$ and $\hat{H}_2$ it will be convenient to define the following states $\bigl\vert\chi\bigr>_{\alpha,\beta}$ by

\begin{eqnarray}
\label{REPRESENT28}
\bigl\vert\chi\bigr>_{\alpha,\beta}\equiv\bigl\vert\chi\bigr>\otimes\bigl\vert\alpha\bigr>\otimes\bigl\vert\beta\bigr>.
\end{eqnarray}

\noindent
We will replace the action of $\hat{a}_1$ and $\hat{a}_2$ on (\ref{REPRESENT28}) by their eigenvalues, and leave the operator $\hat{a}_3$ in its present form since we have singled out $\hat{a}_3$ as special.  Then the following relations ensue

\begin{eqnarray}
\label{IDENTITY}
\hat{H}_1\bigl\vert\chi\bigr>_{\alpha,\beta}
=\Bigl((\hat{a}_3+\gamma^{-})(\hat{a}_3+\gamma^{+})+l\hat{a}_3(\hat{a}_3+\alpha)(\hat{a}_3+\beta)e^{-\hat{a}_3^{\dagger}}\Bigr)\bigl\vert\chi\bigr>_{\alpha,\beta}
\end{eqnarray}

\noindent
and

\begin{eqnarray}
\label{IDENTITY1}
\hat{H}_2\bigl\vert\chi\bigr>_{\alpha,\beta}
=\Bigl(\hat{a}_3(\hat{a}_3+\alpha)(\hat{a}_3+\beta)+r(\hat{a}_3+\gamma^{-})(\hat{a}_3+\gamma^{+})e^{\hat{a}^{\dagger}_3}\Bigr)\bigl\vert\chi\bigr>_{\alpha,\beta}.
\end{eqnarray}

\noindent
Having defined the operators and algebra of our system, we will next associate the system to gravity.  First let us associate to each point $x$ in 3-space $\Sigma$ a harmonic oscillator of the type (\ref{WEHAVE}), as in

\begin{eqnarray}
\label{WEEHAVE}
S=\Bigl{\{}a_1(x),a_2(x),a_3(x),a^{*}_1(x),a^{*}_2(x),a^{*}_3(x),1\Bigr{\}}.
\end{eqnarray}

\noindent
Then all of the aforementioned formalism can be repeated for each $x\in\Sigma$.  If 3-space $\Sigma$ were continuous then we would have an infinite number of representations of the oscillator algebra, one representation per point.  But let us start with the assumption that space is discrete, and then we can always attempt to take the continuum limit of the resulting theory.

\section{Holomorphic Schr\"odinger representation}

Perform a 3+1 decomposition of 4-dimensional spacetime $M=\Sigma\times{R}$ where $\Sigma$ is a 3-dimensional spatial manifold, and define by $\Delta_N(\Sigma)$ a discretization of $\Sigma$ on a lattice 
of spacing $\epsilon={{l^3} \over N}$, where $l$ is some characteristic linear dimension associated to $\Sigma$ and $N$ is the total number of lattice sites.  For each $x\in\Delta_N(\Sigma_t)$ on the final 
spatial hypersurface $\Sigma_t$ labelled by coordinate time $t$ define quantities $(X,Y,T)$, which are elements of the space of holomorphic functions, by

\begin{eqnarray}
\label{STATE}
\bigl(X(x,t),Y(x,t),T(x,t)\bigr)\in\Gamma_{Kin}
\end{eqnarray}

\noindent
where $\Gamma_{Kin}$, as defined in the introduction, is the kinematic configuration space at each point $x$ on the hypersurface $\Sigma_t$.  Also define $\forall{x}\in\Delta_N(\Sigma)$ a two dimensional complex space coordinatized 
by $(\widetilde{\alpha}_x,\widetilde{\beta}_x)\in{C}_2$ and associate with each $C_2(x)$ a state $\chi(T_x(t))\otimes\bigl\vert\widetilde{\alpha},\widetilde{\beta}\bigr>_x$, where

\begin{eqnarray}
\label{STATE1}
\chi(T_x(t))=e^{\nu(\hbar{G})^{-1}\int_{\Gamma}\lambda(T)\delta{T}}
\end{eqnarray}

\noindent
for $\lambda(T)\in{C}^{\infty}(\Gamma_{Kin})$.\footnote{The notation $\int_{\Gamma}$ signifies that the integration must be carried out in functional space of the field $T$.  The integration is defined independently for each 
point $x\in\Sigma_t$ on the spatial hypersurface corresponding to time $t$.}  Hence we assume that the antiderivative in the 
exponential of (\ref{STATE1}) exists.  The following mass dimensions are defined for the various quantities of interest

\begin{eqnarray}
\label{MASSDIM}
[X]=[Y]=[T]=0;~~[\nu]=-3;~~[\widetilde{\lambda}]=[\widetilde{\alpha}]=[\widetilde{\beta}]=1.
\end{eqnarray}

\noindent
Let the state $\bigl\vert\widetilde{\alpha},\widetilde{\beta}\bigr>_x$ have the following Schr\"odinger representation

\begin{eqnarray}
\label{LETTHE}
e^{(\hbar{G})^{-1}\nu(\alpha_xX_x+\beta_yY_y)}.
\end{eqnarray}

\noindent
Let us form the continuum limit of the part of the state dependent on $(X,Y)$ by the direct product of (\ref{LETTHE}) over all $x\in\Sigma$

\begin{eqnarray}
\label{STATE2}
\boldsymbol{\psi}_{\alpha,\beta}[X,Y]=\bigl<X,Y\bigl\vert\widetilde{\alpha},\widetilde{\beta}\bigr>=\hbox{lim}_{\epsilon\rightarrow{0}}\prod_x\bigl<X_x(t),Y_x(t)\bigl\vert\widetilde{\alpha}_x,\widetilde{\beta}_x\bigr>\nonumber\\
=N(\widetilde{\alpha},\widetilde{\beta})e^{(\hbar{G})^{-1}(\widetilde{\alpha}\cdot{X}+\widetilde{\beta}\cdot{Y})}.
\end{eqnarray}

\noindent
In this limit we have $\Delta_N(\Sigma)\rightarrow\Delta_{\infty}(\Sigma)$, and the dot product signifies a Riemannian integral over 3-space, as in\footnote{This can be seen as the result of assigning a volume of $\nu$ to each point in $\Delta_N(\Sigma)$, as in (\ref{LETTHE}).  In the continuum limit the sum over each volume $\nu$ becomes a Riemannian integral.}

\begin{eqnarray}
\label{STATE3}
U\cdot{V}=\int_{\Sigma}d^3xU(x)V(x)~\forall~{U},V\in{C}^0(\Sigma).
\end{eqnarray}

\noindent
The quantity $N(\widetilde{\alpha},\widetilde{\beta})$ in (\ref{STATE2}) is a normalization factor given by

\begin{eqnarray}
\label{STATE4}
N(\widetilde{\alpha},\widetilde{\beta})=e^{-\nu(\hbar{G})^{-2}(\widetilde{\alpha}^{*}\cdot\widetilde{\alpha}+\widetilde{\beta}^{*}\cdot\widetilde{\beta})}.
\end{eqnarray}

\noindent
Note that the states $\boldsymbol{\psi}_{\alpha,\beta}\in{L}_2(\Gamma_{Kin},D\mu)$ are square-integrable with respect to the measure

\begin{eqnarray}
\label{STATE5}
D\mu=\prod_xD(X,Y)_xe^{-\nu^{-1}(\overline{X}\cdot{X}+\overline{Y}\cdot{Y})},
\end{eqnarray}

\noindent
where $D(X,Y)_x=\delta\overline{X}\delta{X}\delta\overline{Y}\delta{Y}$ and $\nu$ is a numerical constant of mass dimension $[\nu]=-3$.  The overlap between two states in the measure (\ref{STATE5}) is given by

\begin{eqnarray}
\label{STATE6}
\bigl\vert\bigl<\widetilde{\alpha},\widetilde{\beta}\bigl\vert\widetilde{\alpha}^{\prime},\widetilde{\beta}^{\prime}\bigr>\bigr\vert^2
=\hbox{exp}\Bigl[-\nu(\hbar{G})^{-2}\int_{\Sigma}d^3x\Bigl(\bigl\vert\widetilde{\alpha}(x)-\widetilde{\alpha}^{\prime}(x)\bigr\vert^2+\bigl\vert\widetilde{\beta}(x)-\widetilde{\beta}^{\prime}(x)\bigr\vert^2\Bigr)\Bigr],
\end{eqnarray}

\noindent
which is inversely proportional to the Euclidean distance between the state labels in the two dimensional complex manifold $C_2$.  Let us first consider a special case where $\lambda(T)$ is independent of $T$, given by

\begin{eqnarray}
\label{STATE7}
\widetilde{\lambda}(T)=\widetilde{\lambda}^{\pm}_{\alpha,\beta}=-{1 \over 3}\bigl(\widetilde{\alpha}+\widetilde{\beta}\pm\sqrt{\widetilde{\alpha}^2-\widetilde{\alpha}\widetilde{\beta}+\widetilde{\beta}^2}\bigr).
\end{eqnarray}

\noindent
In this case (\ref{STATE1}) yields $\chi(T)=e^{(\hbar{G})^{-1}\lambda^{\pm}_{\alpha,\beta}\cdot{T}}$ which produces a state

\begin{eqnarray}
\label{STATE8}
\boldsymbol{\psi}^0_{\alpha,\beta}[X,Y,T]=e^{(\hbar{G})^{-1}(\alpha\cdot{X}+\beta\cdot{Y}+\lambda^{\pm}_{\alpha,\beta}\cdot{T})}.
\end{eqnarray}

\noindent
Define dynamical momentum space variables $\Pi(x,t)$, $\Pi_1(x,t)$ and $\Pi_2(x,t)$ on the kinematic momentum space $P_{Kin}$, which upon quantization become promoted to operators satisying equal-time commutation relations

\begin{eqnarray}
\label{STATE9}
\bigl[\hat{T}(x,t),\hat{\Pi}(y,t)\bigr]=\bigl[\hat{X}(x,t),\hat{\Pi}_1(y,t)\bigr]=\bigl[\hat{Y}(x,t),\hat{\Pi}_2(y,t)\bigr]=(\hbar{G})\delta^{(3)}(x,y).
\end{eqnarray}

\noindent
Also define the following function $Q$ on the kinematic momentum space $P_{Kin}$, given by

\begin{eqnarray}
\label{STATE10}
Q=\Pi^2+{2 \over 3}(\Pi_1+\Pi_2)\Pi+{1 \over 3}\Pi_1\Pi_2.
\end{eqnarray}

\noindent
Equation (\ref{STATE10}) can be written in the equivalent form by dividing it by $\Pi(\Pi+\Pi_1)(\Pi+\Pi_2)\neq{0}$, which yields

\begin{eqnarray}
\label{STATE13}
{1 \over \Pi}+{1 \over {\Pi+\Pi_1}}+{1 \over {\Pi+\Pi_2}}=0.
\end{eqnarray}

\noindent
Note that $\bigl\vert\widetilde{\alpha},\widetilde{\beta}\bigr>$ are eigenstates of $\hat{\Pi}_1$ and $\hat{\Pi}_2$, given in the functional Schr\"odinger representation by

\begin{eqnarray}
\label{STATE11}
\hat{\Pi}(x)\boldsymbol{\psi}=(\hbar{G}){\delta \over {\delta{T}(x)}}\boldsymbol{\psi};\nonumber\\
\hat{\Pi}_1(x)\bigl\vert\widetilde{\alpha}\bigr>\longrightarrow(\hbar{G}){\delta \over {\delta{X}(x)}}e^{(\hbar{G})^{-1}\widetilde{\alpha}\cdot{X}}\longrightarrow\widetilde{\alpha}(x)\bigl\vert\widetilde{\alpha}\bigr>;\nonumber\\
\hat{\Pi}_2(x)\bigl\vert\widetilde{\beta}\bigr>\longrightarrow(\hbar{G}){\delta \over {\delta{Y}(x)}}e^{(\hbar{G})^{-1}\widetilde{\beta}\cdot{Y}}\longrightarrow\widetilde{\beta}(x)\bigl\vert\widetilde{\beta}\bigr>.
\end{eqnarray}

\noindent
Also note that $\boldsymbol{\psi}^0_{\alpha,\beta}\in{Ker}\{\hat{Q}\}$, which can also be written as

\begin{eqnarray}
\label{STATE12}
\Bigl((\hbar{G})^2{{\delta^2} \over {\delta{T}(x)\delta{T}(x)}}+{2 \over 3}(\alpha+\beta)(\hbar{G}){\delta \over {\delta{T}(x)}}+{1 \over 3}\alpha\beta\Bigr)\boldsymbol{\psi}^0_{\alpha,\beta}=0,
\end{eqnarray}

\noindent
where we have replaced the actions of $\Pi_1$ and $\Pi_2$ by their eigenvalues on the state.  We have left the action of $\Pi$ intact as a functional derivative, because we have singled $T(x)$ as a time variable on $\Gamma_{Kin}$ and we will be interested in the evolution of the state with respect to $T$.  Equations (\ref{STATE11}) are the continuum limit of the following discretized versions for $x\in\Delta_N(\Sigma)$

\begin{eqnarray}
\label{STATE111}
\hat{\Pi}_x\boldsymbol{\psi}=(\hbar{G})\nu^{-1}{\partial \over {\partial{T}_x}}\boldsymbol{\psi};\nonumber\\
(\hat{\Pi}_1)_x\bigl\vert\widetilde{\alpha}\bigr>\longrightarrow(\hbar{G})\nu^{-1}{\partial \over {\partial{X}_x}}e^{\nu(\hbar{G})^{-1}\alpha{X}}\longrightarrow\widetilde{\alpha}_x\bigl\vert\widetilde{\alpha}\bigr>;\nonumber\\
(\hat{\Pi}_2)_x=\bigl\vert\widetilde{\beta}\bigr>\longrightarrow(\hbar{G})\nu^{-1}{\partial \over {\partial{Y}_x}}e^{\nu(\hbar{G})^{-1}\beta{Y}}\longrightarrow\widetilde{\beta}_x\bigl\vert\widetilde{\beta}\bigr>,
\end{eqnarray}

\noindent
whence the integration has been restricted to a single cell of characteristic volume dimension $\nu$ containing the point $x$.  The effect of the the factor $\nu^{-1}$ in the partial derivative is the analogue of a delta function in the functional derivative of the continuum limit.  Similarly, the discretized version of (\ref{STATE12}) is given by

\begin{eqnarray}
\label{STATE112}
\Bigl((\hbar{G}\nu^{-1})^2{{\partial^2} \over {\partial{T}_x}}+{2 \over 3}(\widetilde{\alpha}_x+\widetilde{\beta}_x)
(\hbar{G}\nu^{-1}){\partial \over {\partial{T}_x}}+{1 \over 3}\widetilde{\alpha}_x\widetilde{\beta}_x\Bigr)\boldsymbol{\psi}^0_{\alpha,\beta}=0.
\end{eqnarray}

\noindent
We will now make an association from the holomorphic states $\boldsymbol{\psi}^0_{\alpha,\beta}$ constructed in this section to gravity in two stages.  First we will show how the Hilbert space follows from the kinematic level of the instanton representation of Plebanski gravity.  Secondly, we will provide an embedding map from the kinematic phase space to the unconstrained phase space which we will in turn map into the Ashtekar variables.

\section{Transformation into the instanton representation action of Plebanski gravity}

\noindent
We will now construct an action which upon quantization yields the commutation relations (\ref{STATE9}) and the constraint (\ref{STATE13}).  This is given by

\begin{eqnarray}
\label{STATE14}
I_{Kin}={i \over G}\int{dt}\int_{\Sigma}d^3x\Bigl(\Pi\dot{T}+\Pi_1\dot{X}+\Pi_2\dot{Y}\nonumber\\
-iNK\sqrt{\Pi(\Pi+\Pi_1)(\Pi+\Pi_2)}\Bigl({1 \over \Pi}+{1 \over {\Pi+\Pi_1}}+{1 \over {\Pi+\Pi_2}}\Bigr)\Bigr),
\end{eqnarray}

\noindent
where $K=K(X,Y,T)\neq{0}$ is some function of the kinematic configuration space variables $X,Y,T\in\Gamma_{Kin}$, which will be choosen appropriately.  Note that (\ref{STATE14}) implies the symplectic two form

\begin{eqnarray}
\label{STATE15}
\boldsymbol{\omega}_{Kin}={i \over G}\int_{\Sigma}d^3x\Bigl({\delta\Pi}\wedge{\delta{T}}+{\delta\Pi_1}\wedge{\delta{X}}+{\delta\Pi_2}\wedge{\delta{Y}}\Bigr)\nonumber\\
={i \over G}\delta\Bigl(\int_{\Sigma}d^3x\bigl(\Pi\delta{T}+\Pi_1\delta{X}+\Pi_2\delta{Y}\bigr)\Bigr)\equiv\delta\boldsymbol{\theta}_{Kin},
\end{eqnarray}

\noindent
where $\boldsymbol{\theta}_{Kin}$ is the canonical one form on the kinematic phase space $\Omega_{Kin}$.  We will now perform a change of variables.  Define a mass scale $a_0=const.$ and define new momentum 
space variables $(\lambda_1,\lambda_2,\lambda_3)$ such that

\begin{eqnarray}
\label{STATE16}
\Pi_1=a_0^3e^T(\lambda_1-\lambda_3);~~\Pi_2=a_0^3e^T(\lambda_2-\lambda_3);~~\Pi=a_0^3e^T\lambda_3,
\end{eqnarray}

\noindent
and define new configuration space variables $(a_1,a_2,a_3)$ such that\footnote{Note that $a_1$, $a_2$ and $a_3$ are not to be confused with the harmonic oscillator annihilation operators of the previous sections.}

\begin{eqnarray}
\label{STATE17}
a_1=a_0e^X;~~a_2=a_0e^Y;~~a_1a_2a_3=a_0^3e^T.
\end{eqnarray}

\noindent
The ranges of the coordinates are $-\infty<\vert{X}\vert,\vert{Y}\vert,\vert{T}\vert<\infty$ where

\begin{eqnarray}
\label{WHERE}
\vert{a}\vert=\sqrt{(Re\{a\})^2+(Im\{a\})^2},
\end{eqnarray}

\noindent
which corresponds to $0<\vert{a}_f\vert<\infty$.  Under the transformations (\ref{STATE16}) 
and (\ref{STATE17}), then the action (\ref{STATE14}) is given by

\begin{eqnarray}
\label{STATE18}
I_{Kin}={i \over G}\int{dt}\int_{\Sigma}d^4x\Bigl(\lambda_1a_2a_3\dot{a}_1+\lambda_2a_3a_1\dot{a}_2+\lambda_3a_1a_2\dot{a}_3\nonumber\\
-iNK\sqrt{\lambda_1\lambda_2\lambda_3}\Bigl({1 \over {\lambda_1}}+{1 \over {\lambda_2}}+{1 \over {\lambda_3}}\Bigr)\Bigr)
\end{eqnarray}

\noindent
where now $K=K(a_1,a_2,a_3)$, which will be chosen appropriately.  We will now adopt the following convention for indices, where symbols from the beginning of the Latin alphabet $a,b,c\dots$ signify internal indices and symbols from the middle $i,j,k,\dots$ signify spatial indices in $\Sigma$.  We will associate the internal indices with $SO(3,C)$, the special complex orthogonal group in three dimensions.  Let us now make the following identifications

\begin{displaymath}
\alpha^a_i=
\left(\begin{array}{ccc}
a_1 & 0 & 0\\
0 & a_2 & 0\\
0 & 0 & a_3\\
\end{array}\right)
;~~\beta^i_a=\epsilon^{ijk}\partial_j\alpha^a_k+{1 \over 2}\epsilon^{ijk}f^{abc}\alpha^b_j\alpha^c_k,
\end{displaymath}

\noindent
where $\beta^i_a$ will play the role of a magnetic field for $\alpha^a_i$, seen as a nonabelian gauge field.  Note for the diagonal $\alpha^a_i=\delta^a_ia_a$ that there are no spatial gradients in the canonical 
one form $\boldsymbol{\theta}_{Kin}$.\footnote{This is because, due to the antisymmetry of $\epsilon^{ijk}$ and the symmetry of a diagonal connection $\delta^a_ia_i$, that the spatial gradient terms drop out.  Since the spatial gradients are still nonzero, we are dealing with the full theory and not minisuperspace.  There are three degrees of freedom per point in the diagonal connection.}  Let us define a new variable $\Psi_{ae}$, given by

\begin{displaymath}
\Psi_{ae}=(e^{\vec{\theta}\cdot{T}})_{af}
\left(\begin{array}{ccc}
\lambda_1 & 0 & 0\\
0 & \lambda_2 & 0\\
0 & 0 & \lambda_3\\
\end{array}\right)_{fg}
(e^{-\vec{\theta}\cdot{T}})_{ge}+\epsilon_{aed}\psi^d,
\end{displaymath}

\noindent
where $\vec{\theta}=(\theta^1,\theta^2,\theta^3)\in{C}_3$ are a triplet of complex rotation parameters, $T$ correspond to the $SO(3)$ generators and $\psi^d$ is a $SO(3,C)$- valued 3-vector.  Note for $\psi^d=0$ that $\Psi_{ae}$ is symmetric in $a,e$, since it takes on the interpretation of an $SO(3,C)$ transformation of the diagonal matrix of eigenvalues.  Define the following quantities

\begin{eqnarray}
\label{STATE19}
b^i_a=(e^{\vec{\theta}\cdot{T}})_{ae}\beta^i_e;~~a^a_i=(e^{\vec{\theta}\cdot{T}})_{ae}\alpha^e_i+{1 \over 2}\epsilon^{abc}(e^{\vec{\theta}\cdot{T}})_{bf}\partial_i(e^{\vec{\theta}\cdot{T}})_{cf}.
\end{eqnarray}

\noindent
Note that $b^i_a=b^i_a(\vec{a},\vec{\theta})$ is the result of rotating the internal index of $\beta^i_a$, which corresponds a $SO(3,C)$ transformation.  It then follows that $a^a_i=a^a_i(\vec{a},\vec{\theta})$, which now has six degrees of freedom, is the corresponding gauge transformed version of $\alpha^a_i=Diag(a_1,a_2,a_3)$ which has just three degrees of freedom.  The transformation (\ref{STATE19}) induces an 
embedding $\Omega_{Kin}\rightarrow\Omega_{diff}$, where $\Omega_{diff}$ is defined as a diffeomorphism invariant phase space with action

\begin{eqnarray}
\label{STATE20}
I_{diff}={i \over G}\int{dt}\int_{\Sigma}d^3x\Bigl(\Psi_{(ae)}b^i_e\dot{a}^a_i-iN(\hbox{det}b)^{1/2}\sqrt{\hbox{det}\Psi}\hbox{tr}\Psi^{-1}\Bigr)\biggl\vert_{Sym(\Psi)}.
\end{eqnarray}

\noindent
From (\ref{STATE20}) one makes the identification $K\equiv\hbox{det}b$.  By the notation $Sym(\Psi)$ is meant that $\Psi_{ae}=\Psi_{ea}$ is symmetric.  We can remove this restriction by allowing $\Psi$ to have an antisymmetric part while imposing the constraint that this antisymmetric part vanishes.  We can also constrain the $SO(3,C)$ frame by imposing a constraint on $\vec{\theta}$.  In conjunction with the aformentioned constraints and the constraint on the eigenvalues $\lambda_f$ we will impose the following constraints on the unreduced phase space $\Omega_{Inst}$, given by 

\begin{eqnarray}
\label{STATE21}
H=(\hbox{det}b)^{1/2}\sqrt{\hbox{det}\Psi}\hbox{tr}\Psi^{-1}=0;\nonumber\\
H_i=\epsilon_{ijk}b^j_ab^k_e\Psi_{ae}=0;\nonumber\\
G_a=b^i_e\partial_i\Psi_{ae}+\bigl(f_{abf}\delta_{ge}+f_{ebg}\delta_{af}\bigr)a^b_ib^i_e\Psi_{fg}=b^i_eD_i\Psi_{ae}=0.
\end{eqnarray}

\noindent
The constraints (\ref{STATE21}) can be obtained by the variation of Lagrange multipliers $(a^f_0,N,N^i)$ in the following action

\begin{eqnarray}
\label{STATE22}
I_{Inst}={i \over G}\int{dt}\int_{\Sigma}d^3x\Bigl(\Psi_{ae}b^i_e\dot{a}^a_i-a^a_0G_a-N^iH_i-iNH\Bigr).
\end{eqnarray}

\noindent
Note that there is no configuration space variable canonically conjugate to $\Psi_{ae}$, since the canonical one form $\boldsymbol{\theta}=\int_{\Sigma}d^3x\Psi_{ae}b^i_e\delta{a}^a_i$ does not vary into a canonical symplectic two form.\par
\indent
The momentum space $\Psi_{ae}$ of (\ref{STATE22}) has nine degrees of freedom per point, but the connection $a^a_i$ has only six.  We may lift this restriction, in conjunction with lifting the restriction to symmetric $\Psi_{ae}$, and make the identification $a^a_i\rightarrow{A}^a_i$ and $b^i_a\rightarrow{B}^i_a[A]$ where now $A^a_i$ and therefore $B^i_a$ now have nine degrees of freedom per point.  We can then write the extended action as

\begin{eqnarray}
\label{STATE221}
I_{Inst}={i \over G}\int{dt}\int_{\Sigma}d^3x\Bigl(\Psi_{ae}B^i_e\dot{A}^a_i+A^a_0B^i_eD_i\Psi_{ae}\nonumber\\
-\epsilon_{ijk}N^iB^j_aB^k_e\Psi_{ae}-iN(\hbox{det}B)^{1/2}\sqrt{\hbox{det}\Psi}\hbox{tr}\Psi^{-1}\Bigr),
\end{eqnarray}

\noindent
combined with a prescription for obtaining the diffeomorphism invariant phase space $\Omega_{diff}$.  This prescription is to set to zero all components of $A^a_i$ not obtainiable 
from a diagonal connection $Diag(a_1,a_2,a_3)$ by $SO(3,C)$ gauge transformation, in conjunction with setting $\Psi_{[ae]}=0$, when implementing the diffeomorphism constraint $H_i=0$.  Note, in direct analogy 
to (\ref{STATE22}), that $\boldsymbol{\theta}_{Inst}=\int_{\Sigma}d^3x\Psi_{ae}B^i_e\delta{A}^a_i$ also does not yield a canonical symplectic two form.  The phase space variables satisfy 

\begin{eqnarray}
\label{STATE222}
\bigl[A^a_i(x,t),\Psi_{bf}(y,t)\bigr]=(\hbar{G})\delta^a_b(B^{-1})^f_i\delta^{(3)}(x,y),
\end{eqnarray}

\noindent
which are not canonical commutation relations owing to the field dependence on the right hand side.  Note, however, that on the kinematic phase space $\Omega_{Kin}$ 
in (\ref{STATE15}) $\boldsymbol{\omega}_{Kin}=\delta\boldsymbol{\theta}_{Kin}$ which implies canonical commutation relations (\ref{STATE9}).  Equation (\ref{STATE221}) is the action $I_{Inst}$ for Plebanski gravity in the instanton representation for vanishing cosmological constant, derived in \cite{EYO}.  Equation (\ref{STATE14}) is the action on the reduced phase space for gauge transformations and diffeomorphisms, defined as the kinematic 
phase space $\Omega_{Kin}$.

\section{Transformation into the Ashtekar variables}

\noindent
We have performed an embedding map from the kinematic phase space $\Omega_{Kin}$, which has a closed symplectic two form $\omega_{Kin}$, to the unreduced phase space of the instanton 
representation of Plebanski gravity $\Omega_{Inst}$, whose symplectic two form $\omega_{Inst}$ is in general not closed.  But we would like a theory which on its full unconstrained phase space admits a closed symplectic two form, and we would like this theory to admit a well-defined sequence of transformations to $\Omega_{Kin}$ and its resulting Hilbert space.  To deal with this let us make the change of variables

\begin{eqnarray}
\label{STATE23}
\Psi^{-1}_{ae}=B^i_e(\widetilde{\sigma}^{-1})^a_i\biggl\vert_{\hbox{det}\widetilde{\sigma}\neq{0}},
\end{eqnarray}

\noindent
which holds for nondegenerate variables.  Substitution of (\ref{STATE23}) into (\ref{STATE221}) and defining $\underline{N}=N(\hbox{det}\widetilde{\sigma})^{-1/2}$ yields an action

\begin{eqnarray}
\label{STATE24}
I_{Ash}={i \over G}\int{dt}\int_{\Sigma}d^3x\Bigl(\widetilde{\sigma}^i_a\dot{A}^a_i-A^a_0D_i\widetilde{\sigma}^i_a-\epsilon_{ijk}N^i\widetilde{\sigma}^j_aB^k_a
-{i \over 2}\underline{N}\epsilon_{ijk}\epsilon^{abc}\widetilde{\sigma}^i_a\widetilde{\sigma}^j_bB^k_c\Bigr)
\end{eqnarray}

\noindent
with phase space variables $(\widetilde{\sigma}^i_a,A^a_i)$ which upon quantization would satisfy the equal-time canonical commutation relations

\begin{eqnarray}
\label{STATE25}
\bigl[A^a_i(x,t),\widetilde{\sigma}^j_b(y,t)\bigr]=(\hbar{G})\delta^a_b\delta^j_i\delta^{(3)}(x,y).
\end{eqnarray}

\noindent
Note that (\ref{STATE23}) is a noncanonical transformation from $\Omega_{Inst}$ into $\Omega_{Ash}$, the phase space of the Ashtekar variables, where $A^a_i$ is the self-dual Ashtekar connection.  Indeed, (\ref{STATE24}) is the action for general relativity in the Ashtekar variables for vanishing cosmological constant (See e.g. \cite{ASH1}, \cite{ASH2} and \cite{ASH3}).  The symplectic two form corresponding to (\ref{STATE24}) is given by

\begin{eqnarray}
\label{THESYMPLECTIC}
\omega_{Ash}={i \over G}\int_{\Sigma}d^3x{\delta\widetilde{\sigma}^i_a}\wedge{\delta{A^a_i}}={i \over G}\delta\Bigl(\int_{\Sigma}d^3x\widetilde{\sigma}^i_a\delta{A}^a_i\Bigr)=\delta\boldsymbol{\theta}_{Ash},
\end{eqnarray}

\noindent
which is the exact functional variation of the canonical one form $\boldsymbol{\theta}_{Ash}$.\par
\indent

\subsection{Inverse transformation in the case of nonzero $\Lambda$}

Let us now generalize to the case of a nonvanishing cosmological constant $\Lambda$.  The only change to the action (\ref{STATE24}) occurs in the Hamiltonian constraint, which is now given by

\begin{eqnarray}
\label{STATE26}
H=\epsilon_{ijk}\epsilon^{abc}\widetilde{\sigma}^i_a\widetilde{\sigma}^j_bB^k_c+{\Lambda \over 3}\epsilon_{ijk}\epsilon^{abc}\widetilde{\sigma}^i_a\widetilde{\sigma}^j_b\widetilde{\sigma}^k_c.
\end{eqnarray}

\noindent
Performing all of the previous steps from (\ref{STATE}) to (\ref{STATE24}) in reverse to accomplish the projection $\Omega_{Ash}\rightarrow\Omega_{Inst}\rightarrow\Omega_{diff}\rightarrow\Omega_{Kin}$ from the full unconstrained Ashtekar variables to the quantizable kinematic phase space of the instanton representation, we 
find that the analogue of (\ref{STATE13}) for $\Lambda\neq{0}$ is given by

\begin{eqnarray}
\label{STATE27}
{\Lambda \over {a_0^3}}+\Bigl({1 \over \Pi}+{1 \over {\Pi+\Pi_1}}+{1 \over {\Pi+\Pi_2}}\Bigr)e^T=0~\forall~x\in\Delta_N(\Sigma).
\end{eqnarray}

\noindent
The effect of the cosmological constant is to bring a mass scale $\sqrt{\Lambda}$ into the theory.  Equation (\ref{STATE27}) can be written in polynomial form as

\begin{eqnarray}
\label{POLYNOMIALFORM}
\Bigl({\Lambda \over {3a_0^3}}\Bigr)\Pi(\Pi+\Pi_1)(\Pi+\Pi_2)+\Bigl(\Pi^2+{2 \over 3}(\Pi_1+\Pi_2)\Pi+{1 \over 3}\Pi_1\Pi_2\Bigr)e^T=0,
\end{eqnarray}

\noindent
obtained by multiplication by $\Pi(\Pi+\Pi_1)(\Pi+\Pi_2)\neq{0}$.  Upon quantization of (\ref{POLYNOMIALFORM}) we have the following functional differential equation

\begin{eqnarray}
\label{POLYNOMIALFORM1}
\hat{H}\boldsymbol{\psi}=\biggl[(\hbar{G})^3\Bigl({\Lambda \over {3a_0^3}}\Bigr){\delta \over {\delta{T}}}\Bigl({\delta \over {\delta{T}}}+{\delta \over {\delta{X}}}\Bigr)\Bigl({\delta \over {\delta{T}}}+{\delta \over {\delta{Y}}}\Bigr)\nonumber\\
+r(\hbar{G})^2\Bigl({{\delta^2} \over {\delta{T}^2}}+{2 \over 3}\Bigl({\delta \over {\delta{X}}}+{\delta \over {\delta{Y}}}\Bigr){\delta \over {\delta{T}}}
+{1 \over 3}{{\delta^2} \over {\delta{X}\delta{Y}}}\Bigr)e^{T}\biggr]\boldsymbol{\psi}^{\Lambda}_{\alpha,\beta}[T]=0~\forall{x}\in\Sigma,
\end{eqnarray}

\noindent
where $\boldsymbol{\psi}^{\Lambda}_{\alpha,\beta}=\bigl\vert\widetilde{\alpha},\widetilde{\beta}\bigr>\otimes\chi(T)$ is the quantum state.  We can replace the action of the functional derivatives with respect to $X$ and $Y$ on the state with their eigenvalues $\widetilde{\alpha}$ and $\widetilde{\beta}$, yielding

\begin{eqnarray}
\label{POLYNOMIALFORM2}
\hat{H}\boldsymbol{\psi}=\biggl[\Bigl({\Lambda \over {3a_0^3}}\Bigr)(\hbar{G}){\delta \over {\delta{T}}}\Bigl((\hbar{G}){\delta \over {\delta{T}}}+\widetilde{\alpha}\Bigr)
\Bigl((\hbar{G}){\delta \over {\delta{T}}}+\widetilde{\beta}\Bigr)\nonumber\\
+r\Bigl((\hbar{G})^2{{\delta^2} \over {\delta{T}^2}}+{2 \over 3}(\widetilde{\alpha}+\widetilde{\beta}){\delta \over {\delta{T}}}
+{1 \over 3}\widetilde{\alpha}\widetilde{\beta}\Bigr){e}^{T}\biggr]\boldsymbol{\psi}^{\Lambda}_{\alpha,\beta}[T]=0.
\end{eqnarray}

\noindent
Whereas in the $\Lambda=0$ case there was not a problem, one can see that for $\Lambda\neq{0}$ one must deal with the multiple functional derivatives acting at the same point, which can now act on the factor of $e^T$.\footnote{Note that this is not an issue for the $(X,Y)$ dependence, since the action on the state is finite without regularization, which as well highlights the reason why $T$ is special.}  At this point we will perform a 
discretization $\Delta_N(\Sigma)$ of 3-space $\Sigma$.  Then the functional derivatives turn into partial derivatives at a particular point, which are finite.  In this process we must append the inverse characteristic size of a cell in order to preserve the mass 
dimensions as in $\delta/\delta{T}(x)\rightarrow{\nu}^{-1}\partial/\partial{T}_x$, and the Hamiltonian constraint reduces to the following differential equation

\begin{eqnarray}
\label{STATE28}
\Bigl[\mu{\partial \over {\partial{T}}}\Bigl(\mu{\partial \over {\partial{T}}}+\widetilde{\alpha}\Bigr)\Bigl(\mu{\partial \over {\partial{T}}}+\widetilde{\beta}\Bigr)
+\Bigl({{3a_0^3} \over \Lambda}\Bigr)\Bigl(\mu{\partial \over {\partial{T}}}+\widetilde{\lambda}^{-}_{\alpha,\beta}\Bigr)
\Bigl(\mu{\partial \over {\partial{T}}}+\widetilde{\lambda}^{+}_{\alpha,\beta}\Bigr)e^T\Bigr]\chi(T)=0,
\end{eqnarray}

\noindent
where the following quantities are defined

\begin{eqnarray}
\label{DEFINEEED}
\mu={{\hbar{G}} \over \nu};~~\widetilde{\lambda}^{\pm}_{\alpha,\beta}={1 \over 3}\Bigl(\widetilde{\alpha}+\widetilde{\beta}\pm\sqrt{\widetilde{\alpha}^2-\widetilde{\alpha}\widetilde{\beta}+\widetilde{\beta}^2}\Bigr);
~~z\equiv3\Bigl({{a_0^3} \over {\mu\Lambda}}\Bigr)e^T
\end{eqnarray}

\noindent
with mass dimensions $[\mu]=1$ and $[z]=0$.  Additionally we will define the following dimensionless state labels from (\ref{WEHAVETHE}) 

\begin{eqnarray}
\label{DEFINEEED1}
\alpha={{\widetilde{\alpha}} \over \mu};~~\beta={{\widetilde{\beta}} \over \mu};~~\lambda_{\alpha,\beta}={{\widetilde{\lambda}_{\alpha,\beta}} \over \mu},
\end{eqnarray}

\noindent
so that $[\alpha]=[\beta]=[\lambda_{\alpha,\beta}]=0$.  Dividing (\ref{STATE28}) by $\mu^3$ and eliminating $T$ in favor of $z$, we obtain upon commuting the factor of $z$ to the left the following differential equation

\begin{eqnarray}
\label{STATE29}
\Bigl[z{d \over {dz}}\Bigl(z{d \over {dz}}+\alpha\Bigr)\Bigl(z{d \over {dz}}+\beta\Bigr)+z\Bigl(z{d \over {dz}}+\lambda^{-}_{\alpha,\beta}+1\Bigr)
\Bigl(z{d \over {dz}}+\lambda^{+}_{\alpha,\beta}+1\Bigr)\Bigr]\chi(z)=0.
\end{eqnarray}

\noindent
Equation (\ref{STATE29}) is a hypergeometric differential equation with solution

\begin{eqnarray}
\label{STATE30}
\chi(z)={_2F_2}\bigl(\lambda^{-}_{\alpha,\beta}+1,\lambda^{+}_{\alpha,\beta}+1;\alpha+1,\beta+1;z\bigr).
\end{eqnarray}

\noindent
The state is then given by the direct product of these functions over a given discretization of 3-space $\Sigma$

\begin{eqnarray}
\label{STATE31}
\boldsymbol{\Psi}_{\alpha,\beta}=\prod_x\chi(T_x)\bigl\vert\alpha_x,\beta_x\bigr>.
\end{eqnarray}

\noindent
For $\Lambda\neq{0}$ there is a three to one correspondence between states and points in $C_2$, whereas for $\Lambda=0$ there is a two to one correspondence.\footnote{It is shown in \cite{ITAE} that for $\Lambda=0$ the continuum limit in $\Delta_{\infty}(\Sigma)$ exists as part of the same Hilbert space as each discretization $\Delta_N(\Sigma)$, but for $\Lambda=0$ the Kodama state $\boldsymbol{\psi}_{Kod}$ is the only state with this property.  In the latter case the discretized Hilbert space converges to elements $\boldsymbol{\Psi}\not\subset{Ker}\{\hat{H}\}$ in the continuum limit, which requires the inclusion of these elements $\boldsymbol{\Psi}$ to complete the Hilbert space.}  Later in this paper we will make the direct association from $\alpha$ and $\beta$ as defined in (\ref{DEFINEEED1}) to the labels of the harmonic oscillator coherent states derived in section 2.  The associated formalism and results from the holomorphic Schr\"odinger representation carry over directly into the coherent state formalism.

\section{Physical interpretation}

We shall now elucidate upon the relation to general relativity of the Hilbert space constructed in the previous sections.  Perform the following decomposition of $\Psi^{-1}_{ae}$

\begin{eqnarray}
\label{MAGNET}
\Psi^{-1}_{ae}=-{\Lambda \over 3}\delta_{ae}+\psi_{ae},
\end{eqnarray}

\noindent
where $\psi_{ae}$ is symmetric and traceless.  In the language of $SL(2,C)$ $Weyl$, shorthand for the self-dual part of the Weyl curvature tensor, can be written in 
unprimed $SL(2,C)$ indices as

\begin{eqnarray}
\label{WHILE}
\psi_{ABCD}=\psi_{(ABCD)}=\eta_{AB}^a\eta_{CD}^e\psi_{ae},
\end{eqnarray}

\noindent
which is totally symmetric in uppercase indices.  We have $A=0,1$ and $a=1,2,3$, where $\eta^a_{AB}$ is an isomorphism from $SL(2,C)$ unprimed index pairs $AB=(00,01,11)$ to single $SO(3,C)$ indices $a=(1,2,3)$.\par
\indent
The eigenvalues of $\psi_{ae}$ encode the algebraic classification of spacetime \cite{GROUP}, which are independent of coordinates and of tetrad frames \cite{PENROSERIND}.  These properties play a role in the determination of the principal null directions and the radiation properties of spacetime \cite{SWALD},\cite{PETROV}.  These properties can be computed from the characteristic equation for $\psi_{ae}$ and the invariants $(I,J)$, given by

\begin{eqnarray}
\label{SPECIATLY}
I=\psi_{ABCD}\psi^{ABCD};~~J=\psi_{ABCD}\psi^{CD}_{EF}\psi^{EFAB}.
\end{eqnarray}
 
\noindent
To make the link from these properties of spacetime to the degrees of freedom that have been quantized, equation (\ref{MAGNET}) can be inverted.  Since $\psi_{ae}=\psi_{ae}(I,J)$ encodes the classification of the spacetime, it follows that $\Psi_{ae}=\Psi_{ae}(I,J)$ also encodes this classification.\par
\indent
In the intrinsic frame $SO(3,C)$ frame, defined as the frame in which $\Psi_{ae}$ is diagonalized, the eigenvalues are given in terms of the state labels by

\begin{displaymath}
\widetilde{\Psi}_{ae}=\Psi_{ae}a_0^3e^T=
\left(\begin{array}{ccc}
\widetilde{\alpha}+\widetilde{\lambda}_{\alpha,\beta} & 0 & 0\\
0 & \widetilde{\beta}+\widetilde{\lambda}_{\alpha,\beta} & 0\\
0 & 0 & \widetilde{\lambda}_{\alpha,\beta}\\
\end{array}\right)
.
\end{displaymath}

\noindent
The states then imply the following classification scheme for spacetimes\footnote{We have adapted the results of \cite{PENROSERIND}, which refer just to $\psi_{ae}$, in terms of $\Psi_{ae}$.}

\begin{eqnarray}
\label{STATE32}
\alpha=\beta=0:~~Petrov~Type~O~(Kodama~state~\boldsymbol{\psi}_{Kod});\nonumber\\
\alpha=\beta\neq{0}:~~Petrov~Type~D~(Algebraically~special);\nonumber\\
\alpha\neq\beta\neq{0}:~~Petrov~Type~I~(Algebraically~general).
\end{eqnarray}

\noindent
To obtain a physical interpretation into the meaning of the densitized eigenvalues, let us examine them in the original variables

\begin{eqnarray}
\label{ORIGINAL}
\lambda={{\widetilde{\lambda}} \over \mu}=\Bigl({{\lambda\nu} \over {\hbar{G}}}\Bigr)=\Bigl({{a_0^3\nu} \over {\hbar{G}}}\Bigr)\lambda_3e^T.
\end{eqnarray}

\noindent
The state labels depend on the mass space $a_0$ for the connection as well as the volume scale $\nu$ of the elementary cells of the discretization.  Since these have so far remained unspecified, let us fix them by making 
the choice $a_0^3\nu=1$, which sets the mass scale $a_0$ to the inverse length scale $\nu^{1/3}$.  Then we have $\widetilde{\lambda}_{\alpha,\beta}=(\hbar{G})^{-1}\lambda_3e^T$, or that the state labels occur in multiplies of the (undensitized) eigenvalues of the CDJ matrix $\Psi_{ae}$.  Since $\Psi^{-1}_{ae}$ is the self-dual part of the Weyl curvature tensor with a trace added in, then it has the same dimensions 
as curvature which are inverse length squared.  In our case the length scale referred to is the Planck length $l_{Pl}\sim\sqrt{\hbar{G}}$.  Hence $\widetilde{\lambda}_{\alpha,\beta}$ can be seen as of the same order of magnitude of variations of a spacetime metric $g_{\mu\nu}$ on the scale of the Planck length $l_{Pl}$.  With this choice of $a_0$ the Hamiltonian constraint takes on the form

\begin{eqnarray}
\label{TAKESONTHE}
H=\nu\Lambda{e}^{-T}+{1 \over \Pi}+{1 \over {\Pi+\Pi_1}}+{1 \over {\Pi+\Pi_2}}=0,
\end{eqnarray}

\noindent
which as we have shown yields a solution for the states in terms of hypergeometric functions.  In undensitized variables this is given by

\begin{eqnarray}
\label{TAKESONTHE1}
H=\Lambda+{1 \over {\lambda_1}}+{1 \over {\lambda_2}}+{1 \over {\lambda_3}}=0,
\end{eqnarray}

\noindent
which is transparent to the parameters introduced as a result of the quantization process.

\section{Recapitulation: Lippman--Schwinger expansion on coherent state basis}

Let us now expand upon the manifestation of the hypergeometric solutions to the Hamiltonian constraint in terms of the coherent state formalism of section 2, continuing from (\ref{IDENTITY}) and (\ref{IDENTITY1}).  The solution to the Hamiltonian constraint consists of states in the kernel of $\hat{H}_1$ and $\hat{H}_2$.  We will build these states by expansion about $Ker\{\hat{Q}\}$ and $Ker\{\hat{O}\}$.  For the first case we have

\begin{eqnarray}
\label{EXPANSION}
\hat{H}_1\bigl\vert\boldsymbol{\psi}\bigr>_1=\bigl(\hat{Q}+l\hat{O}e^{-a_3^{\dagger}}\bigr)\bigl\vert\boldsymbol{\psi}\bigr>=0.
\end{eqnarray}

\noindent
Now act on both sides of (\ref{EXPANSION}) with $\hat{Q}^{-1}$, yielding

\begin{eqnarray}
\label{EXPANSION1}
\bigl(1+l\hat{Q}^{-1}\hat{O}e^{-a_3^{\dagger}}\bigr)\bigl\vert\boldsymbol{\psi}\bigr>_1=\bigl\vert\alpha,\beta,\lambda_{\alpha,\beta}\bigr>.
\end{eqnarray}

\noindent
where $\bigl\vert\alpha,\beta,\lambda_{\alpha,\beta}\bigr>\in{Ker}\{\hat{Q}\}$.  Acting on (\ref{EXPANSION1}) with the inverse of the operator in brackets, we have

\begin{eqnarray}
\label{EXPANSION2}
\bigl\vert\boldsymbol{\psi}\bigr>_1=
\bigl(1+l\hat{Q}^{-1}\hat{O}e^{-a_3^{\dagger}}\bigr)^{-1}
\bigl\vert\alpha,\beta,\lambda\bigr>
=\sum_{n=0}^{\infty}(-l)^n(\hat{Q}^{-1}\hat{O}e^{-a_3^{\dagger}})^n\bigl\vert\alpha,\beta,\lambda_{\alpha,\beta}\bigr>.
\end{eqnarray}

\noindent
Likewise, for $\bigl\vert\boldsymbol{\psi}\bigr>_2\in{Ker}\{\hat{H}_2\}$ we have

\begin{eqnarray}
\label{EXPANSION3}
\bigl\vert\boldsymbol{\psi}\bigr>_2=
\bigl(1+r\hat{O}^{-1}\hat{Q}e^{a_3^{\dagger}}\bigr)^{-1}
\bigl\vert\alpha,\beta,\lambda\bigr>
=\sum_{n=0}^{\infty}(-r)^n(\hat{O}^{-1}\hat{Q}e^{a_3^{\dagger}})^n\bigl\vert\alpha,\beta,\lambda\bigr>.
\end{eqnarray}

\noindent
In (\ref{EXPANSION2}) and (\ref{EXPANSION3}), the states are eigenstates of all operators except for the action due to $\hat{a}_3$, which is given by

\begin{eqnarray}
\label{EXPANSION31}
e^{-\hat{a}^{\dagger}_3}\bigl\vert\lambda\bigr>=\bigl\vert\lambda-1\bigr>;~~
e^{\hat{a}^{\dagger}_3}\bigl\vert\lambda\bigr>=\bigl\vert\lambda+1\bigr>.
\end{eqnarray}

\noindent
This induces a raising and lowering action with respect to the $\lambda$ dependence of the state.  Using the representation theory of the harmonic oscillator thus described, (\ref{EXPANSION2}) can be written as

\begin{eqnarray}
\label{EXPANSION4}
\bigl\vert\boldsymbol{\psi}\bigr>_1=\sum_{n=0}^{\infty}(-l)^n\Bigl({{(\alpha+1)_n(\beta+1)_n(\lambda_{\alpha,\beta}+1)_n} \over {(\gamma^{-}+1)_n(\gamma^{-}+1)_n}}\Bigr)\bigl\vert\alpha,\beta,\lambda_{\alpha,\beta}-n\mu\bigr>.
\end{eqnarray}

\noindent
Equation (\ref{EXPANSION4}) is an infinite series with a zero radius of convergence unless we require the series to terminate at finite order.  This leads to the restrictions $\alpha=N$, $\beta=N$ or $\lambda_{\alpha,\beta}=N$ for some integer $N$, which produces an infinite tower of states labelled by $\alpha$ and $N$, as shown in \cite{ITAE}.  For the other states we have that

\begin{eqnarray}
\label{EXPANSION5}
\bigl\vert\boldsymbol{\psi}\bigr>_2=\sum_{n=0}^{\infty}(-r)^n\Bigl({{(\gamma^{-}+1)_n(\gamma^{-}+1)_n} \over {(\alpha+1)_n(\beta+1)_n(\lambda+1)_n}}\Bigr)\bigl\vert\alpha,\beta,\lambda+n\mu\bigr>,
\end{eqnarray}

\noindent
which is convergent without any restrictions on $\alpha$ and $\beta$.

\subsection{Association to quantum gravity}

\noindent
We will now provide the link from the coherent state formalism to the gravity, which follows from the holomorphic Schr\"odinger representation.  Now that we have constructed states in the kernel of the Hamiltonian constraints $H_1$ and $H_2$, we will now transform the constraints and the corresponding states into the Schr\"odinger representation.  First make the following associations 

\begin{eqnarray}
\label{REPRESENT24}
\hat{a}_1\equiv{\delta \over {\delta{X}}};~~\hat{a}_2\equiv{\delta \over {\delta{Y}}};~~\hat{a}_3\equiv{\delta \over {\delta{T}}},
\end{eqnarray}

\noindent
where $X$, $Y$ and $T$ are holomorphic variables.  Hence any arbitrary function $f=f(X,Y,Z)$ is a holomorphic function.  Note that the adjoints of (\ref{REPRESENT24}) have a representation

\begin{eqnarray}
\label{REPRESENT25}
a_1^{\dagger}\equiv{X};~~a_2^{\dagger}\equiv{Y};~~a_3^{\dagger}\equiv{T},
\end{eqnarray}

\noindent
which fixes the measure for normalization essentially as (\ref{STATE5}).  The harmonic oscillator coherent states then have a representation

\begin{eqnarray}
\label{HAVEA}
\boldsymbol{\psi}(X,Y,T)=\bigl<\alpha,\beta,\lambda\bigl\vert{X},Y,Z\bigr>=  e^{\alpha{X}+\beta{Y}+\lambda{T}},
\end{eqnarray}

\noindent
which are normalizable with respect to the Gaussian measure.\par
\indent
Making the identifications (\ref{REPRESENT24}) and (\ref{REPRESENT25}) in $\hat{H}_1$ and $\hat{H}_2$ of (\ref{DEFINING2}), we can transform the Hamiltonian constraints from the oscillator representation into the 
holomorphic Schr\"odinger representation as 

\begin{eqnarray}
\label{REPRESENT26}
\hat{H}_1={{\delta^2} \over {\delta{T}^2}}}+{2 \over 3}\Bigl({{\delta \over {\delta{X}}}+{\delta \over {\delta{Y}}}\Bigr){\delta \over {\delta{T}}}
+{1 \over 3}{{\delta^2} \over {\delta{X}\delta{Y}}}
+l{\delta \over {\delta{T}}}\Bigl({\delta \over {\delta{T}}}+{\delta \over {\delta{X}}}\Bigr)\Bigl({\delta \over {\delta{T}}}+{\delta \over {\delta{Y}}}\Bigr)e^{-T}
\end{eqnarray}

\noindent
and

\begin{eqnarray}
\label{REPRESENT27}
\hat{H}_2=
{\delta \over {\delta{T}}}\Bigl({\delta \over {\delta{T}}}+{\delta \over {\delta{X}}}\Bigr)\Bigl({\delta \over {\delta{T}}}+{\delta \over {\delta{Y}}}\Bigr)
+r\Bigl({{\delta^2} \over {\delta{T}^2}}}+{2 \over 3}\Bigl({{\delta \over {\delta{X}}}+{\delta \over {\delta{Y}}}\Bigr){\delta \over {\delta{T}}}.
+{1 \over 3}{{\delta^2} \over {\delta{X}\delta{Y}}}\Bigr)e^{T}.
\end{eqnarray}

\noindent
The reason that $a_3$ is special in relation to $a_1$ and $a_2$ in (\ref{DEFINING2}) is the same reason that $T$ is special in relation to $X$ and $Y$ in (\ref{REPRESENT26}) and (\ref{REPRESENT27}).  The Hamiltonian constraint 
operators contain $e^{\pm{a}_3^{\dagger}}$, whose action causes a shift in the $\lambda$ dependence of the state by discrete steps.  However, since there is no occurrence of $a_1^{\dagger}$ or of $a_2^{\dagger}$, then the state labels $(\alpha,\beta)$ remain intact under the Hamiltonian action.\footnote{They are in this sense time-independent observables, if one adopts the physical 
interpretation of $a_3^{\dagger}\sim{T}$ as a time variable on the kinematic configuration space $\Gamma_{Kin}$.}  Therefore we may replace the action of $\hat{a}_1$ and $\hat{a}_2$ on the coherent states with their 
eigenvalues $\alpha$ and $\beta$, and focus solely on the dynamics with respect to $T$.\par
\indent
We will use the following notation for the states 

\begin{eqnarray}
\label{NOTATION}
\boldsymbol{\psi}_{\alpha,\beta}^{j}[T]\equiv\bigl\vert\alpha,\beta\bigr>\otimes\chi(T).
\end{eqnarray}

\noindent
The label $j$ will be used to denote multiple states for the same $\alpha,\beta$.\footnote{For vanishing cosmological constant $\Lambda=0$ $j$ will take on the values $1$ and $2$, and for $\Lambda\neq{0}$ it will have three possible values.}

Note in (\ref{IDENTITY}) and (\ref{IDENTITY1}) that the state $\bigl\vert\alpha\bigr>\otimes\bigl\vert\beta\bigr>$ can be omitted, leaving the following differential equation for $\bigl\vert\chi\bigr>$

\begin{eqnarray}
\label{REPRESENT29}
\hat{H}_1\chi_1=\biggl[\Bigl({\delta \over {\delta{T}}}+\gamma^{-}\Bigr)\Bigl({\delta \over {\delta{T}}}+\gamma^{+}\Bigr) 
+l{\delta \over {\delta{T}}}\Bigl({\delta \over {\delta{T}}}+\alpha\Bigr)\Bigl({\delta \over {\delta{T}}}+\beta\Bigr)e^{-T}\biggr]\chi[T]=0
\end{eqnarray}

\noindent
and

\begin{eqnarray}
\label{REPRESENT30}
\hat{H}_2\chi_2=\biggl[{\delta \over {\delta{T}}}\Bigl({\delta \over {\delta{T}}}+\alpha\Bigr)\Bigl({\delta \over {\delta{T}}}+\beta\Bigr)
+r\Bigl({\delta \over {\delta{T}}}+\gamma^{-}\Bigr)\Bigl({\delta \over {\delta{T}}}+\gamma^{+}\Bigr)e^{T}\biggr]\chi[T]=0.
\end{eqnarray}

\noindent
Equations (\ref{REPRESENT29}) and (\ref{REPRESENT30}) are hypergeometric differential equations, with solution

\begin{eqnarray}
\label{REPRESENT31}
\chi_1={_{3}F_2}\bigl(\alpha-1,\beta-1,\lambda-1;\gamma^{-}+1,\gamma^{+}+1;(-le^{-T})\bigr);\nonumber\\
\chi_2={_{2}F_2}\bigl(\gamma^{-}+1,\gamma^{+}+1;\alpha,\beta,\lambda;(-re^T)\bigr).
\end{eqnarray}

\noindent
Let us now make the following identification

\begin{eqnarray}
\label{REPRESENT32}
r={{3a_0^3\nu} \over {\hbar{G}\Lambda}},
\end{eqnarray}

\noindent
where $\Lambda$ is the cosmological constant, and $\nu$ and $a_0$ are numerical constants of mass dimensions $[\nu]=-3$ and $[a_0]=1$.  Then for $\alpha=\beta=0$ $\chi_2$ yields

\begin{eqnarray}
\label{REPRESENT33}
\boldsymbol{\psi}^j_{0,0}=\hbox{exp}\bigl[-3(\hbar{G}\Lambda)^{-1}\nu{a}_0^3e^T\bigr]\equiv\boldsymbol{\psi}_{Kod}.
\end{eqnarray}

\noindent
If we make the identifications

\begin{eqnarray}
\label{REPRESENT34}
X\equiv\hbox{ln}\Bigl({{A^1_1} \over {a_0}}\Bigr);~~Y\equiv\hbox{ln}\Bigl({{A^2_2} \over {a_0}}\Bigr);~~T\equiv\hbox{ln}\Bigl({{A^1_1A^2_2A^3_3} \over {a_0^3}}\Bigr),
\end{eqnarray}

\noindent
then one realizes that (\ref{REPRESENT33}) is the Chern--Simons functional of a diagonal connection, and is nothing more than the Kodama state.  The general state is given by

\begin{eqnarray}
\label{THEGENERAL}
\boldsymbol{\psi}^j_{\alpha,\beta}=e^{\alpha{X}+\beta{Y}}{_{3}F_2}\bigl(\alpha-1,\beta-1,\lambda-1;\gamma^{-}+1,\gamma^{+}+1;(-le^{-T})\bigr);\nonumber\\
\boldsymbol{\psi}^j_{\alpha,\beta}=e^{\alpha{X}+\beta{Y}}{_{2}F_2}\bigl(\gamma^{-}+1,\gamma^{+}+1;\alpha,\beta,\lambda;(-re^T)\bigr),
\end{eqnarray}

\noindent
which are labelled by two arbitrary parameters.  If we repeat the same construction at each point in 3-space $\Sigma$ as in \cite{ITAE}, then we obtain the functionals

\begin{eqnarray}
\label{THEGENERALONE}
\boldsymbol{\Psi}^j_{\alpha,\beta}=\prod_xe^{\alpha\cdot{X}+\beta\cdot{Y}}{_{3}F_2}\bigl(\alpha-1,\beta-1,\lambda-1;\gamma^{-}+1,\gamma^{+}+1;(-le^{-T(x)})\bigr);\nonumber\\
\boldsymbol{\Psi}^j_{\alpha,\beta}=\prod_xe^{\alpha\cdot{X}+\beta\cdot{Y}}{_{2}F_2}\bigl(\gamma^{-}+1,\gamma^{+}+1;\alpha,\beta,\lambda;(-re^T(x))\bigr).
\end{eqnarray}

\noindent
The wavefunctionals (\ref{THEGENERALONE}) correspond to the quantization of the algebraic classification of spacetime as encoded in the $Weyl$, the self-dual part of the Weyl curvature.\footnote{The physical interpretation of the state labels and the canonical structure is treated in detail in \cite{ITAE1}.}  These states are literally gravitational coherent states, since their coherent nature is preserved under evolution in $T$.  The states are expected to exhibit a well-defined semiclassical limit corresponding to the algebraic classification of the spacetimes that they encode.\par
\indent

\section{Summary and discussion}

In reflection upon and comparison with the summary of LQG provided in the introduction, this paper has accomplished the following tasks.  From a kinematic Hilbert space $H_{Kin}$, we have constructed a physical Hilbert 
space of states $H_{Phys}$ annihilated by the Hamiltonian constraint for gravity.  The states in $H_{Kin}$ are eigenstates of the densitized eigenvalues of the Weyl curvature tensor, which carry two labels per point $(\alpha,\beta)$, and form an overcomplete basis.  These states, harmonic oscillator-like coherent states, are defined at each point in 3-space $\Sigma$ in contrast to the spin network states of LQG, which are defined only on one-dimensional graphs embedded in $\Sigma$.  Our states encode the algebraic classification of a classical geometries corresponding to spacetimes of Petrov Type I, D and O, upon which the coherent states are peaked, and satisfy the quantum Hamiltonian constraint by construction.  A natural time variable $T$ (or alternatively $a_3^{\dagger}$ in oscillator language) emerges on configuration space, with respect to which the state evolves.  This evolution is holomorphic plane-wavelike for $\Lambda=0$ and hypergeometric for $\Lambda\neq{0}$.  If one adopts the interpretation of the $\bigl\vert\alpha,\beta\bigr>$ states as being analogous to the spin network states, then the incorporation of the $T$ dependence would bear the corresponding analogy to the aim of spin foams.\par
\indent  
The results of this paper are limited to reduction from the kinematic level of gravity, which comprises three configuration and three momentum space degrees of freedom with only the Hamiltonian constraint remaining.  We have provided a map from the kinematic phase space $\Omega_{Kin}$ to the larger gravitational phase space of the instanton representation in conjunction with appending the constraints necessary to restore $\Omega_{Kin}$ in congruity with the theory.  Note that the quantization procedure of this paper has been defined only on $\Omega_{Kin}$, and therefore is not presently set up to incorporate the unphysical degrees of freedom of gravity.  So we have in effect applied a reduced phase space quantization with respect to the Gauss' law and diffeomorphism constraints, but a Dirac quantization \cite{DIR} with respect to the Hamiltonian constraint.  This is one aspect differing from LQG, where one has a rigorous procedure in place for dealing with the kinematic constraints of GR at the quantum level.  But nevertheless we have provided a map from the instanton representation to the Ashtekar variables.  The implication of reversal of this and the preceding maps is that starting from the full Ashtekar theory, one has a prescription for reducing the theory to quantizable configurations, and then constructing the corresponding Hilbert space with a well-defined semiclassical limit using coherent states.  Finally, to solidify the link from Ashtekar's gravity to the coherent states, we have put in place the adjointness relations linking the oscillator and the Schr\"odinger formalisms.\par
\indent

\end{document}